\documentclass
[superscriptaddress,secnumarabic,amssymb,amsmath,nobibnotes,aps,prd,showkeys,showpacs,nofootinbib,onecolumn,12pt]{revtex4}%
\usepackage{graphicx}
\usepackage{subfigure}
\usepackage{epsf}
\usepackage{bm}
\usepackage{amsmath}
\usepackage{amsfonts}
\usepackage{amssymb}%
\providecommand{\U}[1]{\protect\rule{.1in}{.1in}}

\newcommand{\be}{\begin{equation}}
\newcommand{\ee}{\end{equation}}

\newcommand{\mincir}{\raise
-3.truept\hbox{\rlap{\hbox{$\sim$}}\raise4.truept\hbox{$<$}\ }}
\newcommand{\magcir}{\raise
-3.truept\hbox{\rlap{\hbox{$\sim$}}\raise4.truept\hbox{$>$}\ }}

\begin{document}
\title{New exact and analytic solutions in Weyl Integrable cosmology from Noether
symmetry analysis}
\author{Andronikos Paliathanasis}
\email{anpaliat@phys.uoa.gr}
\affiliation{Institute of Systems Science, Durban University of Technology, Durban 4000,
South Africa}
\affiliation{Instituto de Ciencias F\'{\i}sicas y Matem\'{a}ticas, Universidad Austral de
Chile, Valdivia 5090000, Chile}

\begin{abstract}
We consider a cosmological model in a
Friedmann--Lema\^{\i}tre--Robertson--Walker background space with an ideal gas
defined in Weyl Integrable gravity. In the Einstein-Weyl theory a scalar field
is introduced in a geometric way. Furthermore, the scalar field and the ideal
gas interact in the gravitational Action Integral. Furthermore, we introduce a
potential term for the scalar field potential and we show that the field
equations admit a minisuperspace description. Noether's theorem is applied for
the constraint of the potential function and the corresponding conservation
laws are constructed. Finally, we solve the Hamilton-Jacobi equation for the
cosmological model and we derive a family of new solutions in Weyl Integrable
cosmology. Some closed-form expressions for the Hubble function are presented.

\end{abstract}
\date{\today}
\maketitle

\section{Introduction}

\label{sec1}

Noether symmetry analysis is a powerful method for the construction of
analytic solutions for dynamical systems which follow from a variational
principle. Noether symmetries have been widely applied in gravitational
physics and cosmology for the investigation of the integrability properties
for the gravitational field equations \cite{ns1,ns2,ns3,ns4,ns5,ns6,ns7}.
Additionally, Noether symmetries have been used as a geometric selection rule
\cite{bas1} for the determination of the unknown parameters and functions in
the gravitational Action Integral ,which have been introduced in order to
explain the cosmological observations \cite{pl1,pl2,pl3,pl4}.

In large scales the universe is observed to be in an acceleration phase, which
it is attributed to an matter source with negative pressure known as dark
energy. For the physical origin of the dark energy various approaches have
been proposed over the last decades in the literature, see for instance
\cite{Ratra,Barrow93,Copeland,Overduin,Basil,Buda,fq,Ferraro,faraonibook,Ame10,rev,ind}%
. In the majority of the dark energy models new degrees of freedom are
introduced in the gravitational Action integral such that the cosmological
dynamics to change in a way in order the physical parameters describe the
observable universe.

In this piece of work, we are interested in the construction of exact and
analytic solutions for the cosmological field equations in Weyl Integrable
Gravity (WIG) also known as Weyl Integrable Spacetime (WIST)
\cite{salim96,ww1,ww2,ww3,ww4,ww5}. In Einstein theory of gravity the
fundamental geometric object of the theory is the metric tensor. On the
contrary, in Weyl-Einstein theory of gravity (in WIG) the fundamental
geometric objects are the metric tensor and a scalar field. The novelty of the
WIG is that the new degrees of freedom are introduced by the geometry of the
physical space. Moreover, because of the geometric characteristics of the
theory, the scalar field of WIG interacts in the Action Integral with the rest
of the matter source of the universe. Cosmological models with interaction in
the dark sector of the universe have been studied before theoretically and
phenomenologically with very interesting results, see for instance
\cite{in2,in4,in5,gr} and references therein .

In our analysis we will show that when the matter source is described by an
ideal gas, for instance by a radiation or a dust fluid source, the
cosmological field equations\ in WIG admit a minisuperspace description and
there exists a Lagrangian function which produces the field equations under
variation. The Noether symmetry analysis is applied for the determination of
the scalar field potential and the derivation of invariant functions and
conservation laws. The plan of the paper is as follows.

In Section \ref{secns} we present the basic properties and definitions for the
theory of variational symmetries, in particular we present the two theorems of
E. Noether for the invariant transformations of the Action Integral. In
Section \ref{sec2} we define our model which is that of a spatially flat
Friedmann--Lema\^{\i}tre--Robertson--Walker (FLRW) universe with an ideal gas
as a matter source defined in WIG. Moreover we determine the minisuperspace
for the cosmological model of our consideration and we write the point-like
Lagrangian which describes the dynamical system.

Furthermore, in Section \ref{sec3}, we apply the Noether symmetry condition in
order to constrain the free function of the cosmological model. We find that
for a specific family of the exponential potential the field equations form a
Hamiltonian integrable system where the Hamilton-Jacobi equation can be solved
in a closed-form expression. Additionally, exact and analytic closed-form
expressions which solve the field equations are presented. Finally, in Section
\ref{sec5} we summarize our results.

\section{Noether symmetry analysis}

\label{secns}

The symmetry analysis is a systematic mathematical method for the
determination of invariant functions, similarity transformations and
conservation laws, for the study of differential equations. Symmetry analysis
was established by S. Lie at the end of the 19th century in a series of books
\cite{lie1,lie2,lie3}. As it was shown by S. Lie, the transformation group
which leaves invariant a given differential equation, or a system of
differential equations, can be used to simplify the differential equation.

E. Noether in 1918 published\ a pioneer work \cite{noe18} for the
determination of conservation laws of differential equations which follow from
a variational principle. Inspired by the spirit of the work of S. Lie, Noether
proved two novel theorems. The first theorem treats the invariance of the
Action Integral under an infinitesimal transformation, while the second
theorem provides a one-to-one correspondence for the symmetries of the Action
Integral with conservation laws for the differential equations. Some similar
studies on finite groups before the work of Noether, are that of Hamel
\cite{h1,h2}, Herglotz \cite{h3}, Kneser \cite{h5} and Klein \cite{h6}, For a
modern discussion on Noether's work we refer the reader to the review
\cite{nonle}.

We continue our discussion of the case of Lagrangian functions which describe
second-order differential equations and invariant transformations with point
symmetries as generators.

Consider the infinitesimal transformation
\begin{align}
t^{\prime}  &  =t+\varepsilon\xi\left(  t,x^{k}\right)  ,\label{tr.01}\\
x^{\prime i}  &  =x^{i}+\varepsilon\eta^{i}\left(  t,x^{k}\right)  ,
\label{tr.02}%
\end{align}
with generator the vector field
\[
X=\frac{\partial t^{\prime}}{\partial\varepsilon}\partial_{t}+\frac{\partial
x^{\prime i}}{\partial\varepsilon}\partial_{i}.
\]
A dot indicates derivative with respect to the variable $t$. For the dynamical
system of second-order ordinary differential equations $H\left(  t,x^{k}%
,\dot{x}^{k},\ddot{x}^{k}\right)  $ which follows from the variation of the
Lagrangian function $\mathcal{L}=\mathcal{L}\left(  t,x^{k},\dot{x}%
^{k}\right)  $, the vector field $X$ is a variational symmetry, i.e. Noether
symmetry, if there exists a function $f$ such that the following condition is
held%
\begin{equation}
X^{\left[  1\right]  }\mathcal{L}+\mathcal{L}\dot{\xi}-\dot{f}=0~.
\label{tr.05}%
\end{equation}

Hence, the Euler-Lagrangian equations, that is the dynamical system $H\left(
t,x^{k},\dot{x}^{k},\ddot{x}^{k}\right)  $ remain invariant under the action
of the point transformation (\ref{tr.01}), (\ref{tr.02}). The symmetry
condition (\ref{tr.05}) is known as Noether's first theorem.

For a given Lagrangian function, equation (\ref{tr.05}) gives a monomial
expression which is identical zero if the coefficients of the independent
monomials are zero. From the latter a system of linear partial differential
equations is defined.

The second Noether's Theorem relates the existence of Noether symmetries to
that of conservation laws. Indeed, if $X$ is the generator of the
infinitesimal transformation (\ref{tr.01}), (\ref{tr.02}) which satisfies the
symmetry condition (\ref{tr.05}) for a specific function$~f$, then the
function%
\begin{equation}
I\left(  t,x^{k},\dot{x}^{k}\right)  =\xi\left(  \frac{\partial\mathcal{L}%
}{\partial\dot{x}^{k}}\dot{x}^{k}-\mathcal{L}\right)  -\eta^{i}\frac
{\partial\mathcal{L}}{\partial\dot{x}^{i}}+f, \label{tr.07}%
\end{equation}
is a conservation law for the dynamical system $H\left(  t,x^{k},\dot{x}%
^{k},\ddot{x}^{k}\right)  $, that is, $\dot{I}\left(  t,x^{k},\dot{x}%
^{k}\right)  =0.$

For a recent review of the application of Noether symmetries in gravitational
theories we refer the reader to \cite{anrv} while other applications of
symmetries in gravitation physics can be found for instance in
\cite{ap1,ap2,ap3,ap4}

\section{Weyl Integrable Gravity}

\label{sec2}

In WIG the geometry is defined by the metric tensor $g_{\mu\nu}$ and the
scalar field $\phi$, in which $\tilde{g}_{\mu\nu}=\phi g_{\mu\nu}.$ The
gravitational Action Integral is considered to be the%
\begin{equation}
S_{W}=\int dx^{4}\sqrt{-g}\left(  \tilde{R}+\xi\left(  \tilde{\nabla}_{\nu
}\left(  \tilde{\nabla}_{\mu}\phi\right)  \right)  g^{\mu\nu}-V\left(
\phi\right)  +L_{m}\right)  , \label{ww.00}%
\end{equation}
in which $\tilde{R}$ is the Ricci scalar of the metric tensor $\tilde{g}%
_{\mu\nu}$, $\xi$ is the coupling constant of the scalar field $\phi$,
$V\left(  \phi\right)  $ is the scalar field potential and $L_{m}$ is the
Lagrangian component which attributes the matter source. In the following we
shall assume that $L_{m}$ describes a perfect fluid.

The covariant derivative $\tilde{\nabla}_{\mu}$ is defined according to the
Christoffel symbols of the metric tensor $\tilde{g}_{\mu\nu}$.

By definition, the following relations for the geometric elements of the
conformal related metrics $\tilde{g}_{\mu\nu}$, $g_{\mu\nu}$ hold,
\begin{equation}
\tilde{\Gamma}_{\mu\nu}^{\kappa}\left(  \tilde{g}\right)  =\Gamma_{\mu\nu
}^{\kappa}\left(  g\right)  -\phi_{,(\mu}\delta_{\nu)}^{\kappa}+\frac{1}%
{2}\phi^{,\kappa}g_{\mu\nu}.
\end{equation}%
\begin{align}
\tilde{R}_{\mu\nu}\left(  \tilde{g}\right)   &  =R_{\mu\nu}\left(  g\right)
-\tilde{\nabla}_{\nu}\left(  \tilde{\nabla}_{\mu}\phi\right)  -\frac{1}%
{2}\left(  \tilde{\nabla}_{\mu}\phi\right)  \left(  \tilde{\nabla}_{\nu}%
\phi\right) \label{ww.04}\\
&  -\frac{1}{2}g_{\mu\nu}\left(  \frac{1}{\sqrt{-g}}\tilde{\nabla}_{\nu}%
\tilde{\nabla}_{\mu}\left(  g^{\mu\nu}\sqrt{-g}\phi\right)  -g^{\mu\nu}\left(
\tilde{\nabla}_{\mu}\phi\right)  \left(  \tilde{\nabla}_{\nu}\phi\right)
\right)  ,\nonumber
\end{align}%
\begin{equation}
\tilde{R}\left(  \tilde{g}\right)  =R\left(  g\right)  -\frac{3}{\sqrt{-g}%
}\tilde{\nabla}_{\nu}\tilde{\nabla}_{\mu}\left(  g^{\mu\nu}\sqrt{-g}%
\phi\right)  +\frac{3}{2}\left(  \tilde{\nabla}_{\mu}\phi\right)  \left(
\tilde{\nabla}_{\nu}\phi\right)  . \label{ww.05}%
\end{equation}

From the gravitational Action Integral (\ref{ww.00}) we derive the
gravitational field equations of Weyl-Einstein theory \cite{salim96}
\begin{equation}
\tilde{G}_{\mu\nu}+\tilde{\nabla}_{\nu}\left(  \tilde{\nabla}_{\mu}%
\phi\right)  -\left(  2\xi-1\right)  \left(  \tilde{\nabla}_{\mu}\phi\right)
\left(  \tilde{\nabla}_{\nu}\phi\right)  +\xi g_{\mu\nu}g^{\kappa\lambda
}\left(  \tilde{\nabla}_{\kappa}\phi\right)  \left(  \tilde{\nabla}_{\lambda
}\phi\right)  -V\left(  \phi\right)  g_{\mu\nu}=e^{-\frac{\phi}{2}}T_{\mu\nu
}^{\left(  m\right)  }, \label{ww.08}%
\end{equation}
where $\tilde{G}_{\mu\nu}\left(  g\right)  $ is the Weyl Einstein tensor and
$T_{\mu\nu}^{\left(  m\right)  }=\left(  \rho_{m}+p_{m}\right)  u_{\mu}u_{\nu
}+p_{m}g_{\mu\nu}$ is the energy momentum tensor which describes the matter
source, $\rho_{m}$, $p_{m}$ are the energy density and pressure components
respectively, while $u^{\mu}$ is the observer.

The field equation (\ref{ww.08}), with the use of (\ref{ww.04}) and
(\ref{ww.05}) can be expressed in the equivalent form
\begin{equation}
G_{\mu\nu}-\lambda\left(  \left(  \nabla_{\mu}\phi\right)  \left(  \nabla
_{\nu}\phi\right)  -\frac{1}{2}g^{\mu\nu}g^{\kappa\lambda}\left(
\nabla_{\kappa}\phi\right)  \left(  \nabla_{\lambda}\phi\right)  \right)
-V\left(  \phi\right)  g_{\mu\nu}=e^{-\frac{\phi}{2}}T_{\mu\nu}^{\left(
m\right)  }, \label{ww.14}%
\end{equation}
in which the new parameter $\lambda$ is defined as $2\lambda\equiv4\xi-3$.
When $\lambda>0$, that is $\xi>\frac{3}{4}$, the scalar field is a real field,
while when $\lambda<0$, i.e. $\xi<\frac{3}{4}$, the scalar field $\phi$ is a
phantom field because its energy density can be negative.

The equations of motion for the scalar field and the fluid source are
\begin{equation}
-g^{\mu\nu}\nabla_{\nu}\nabla_{\mu}\phi+V\left(  \phi\right)  +\frac
{1}{2\lambda}e^{-\frac{\phi}{2}}\rho_{m}=0 \label{ww.15}%
\end{equation}%
\begin{equation}
\left(  \nabla_{\mu}e^{-\phi}\rho_{m}\right)  u^{\mu}+e^{-\phi}\nabla_{\mu
}u^{\mu}\left(  \rho_{m}+p_{m}\right)  =0.
\end{equation}
From the latter expressions it is obvious that the coupling constant $\lambda$
determines the energy transfer and the nature of the interaction between the
two different matter components; the scalar field $\phi$ and the perfect fluid
$\rho_{m}$. For $\lambda>0$ energy transfers from the scalar field to the
perfect fluid, while for $\lambda<0$ the inverse it holds. An interesting
discussion on the nature of the coupling constant and its effects on the
perturbations is presented in \cite{ss1}.

In the following we assume the perfect fluid to be an ideal gas, that is,
$p_{m}=w_{m}\rho_{m}$, in which $w_{m}$ is a constant parameter. For $w_{m}%
=0$, the matter source $\rho_{m}$ describes a pressureless fluid source known
as a dust fluid, while for $w_{m}=\frac{1}{3}$, the matter source $\rho_{m}$
describes a radiation fluid. For this analysis $w_{m}$ is bounded in the
region $w_{m}\in\left(  -1,1\right)  $.

\subsection{FLRW spacetime}

According to the cosmological principle in large scales the universe is
assumed to be isotropic and homogeneous described by the FLRW geometry. In
addition, from the cosmological observations the spatial curvature is very
small, thus the universe is described by the spatially flat FLRW spacetime
with line element%
\begin{equation}
ds^{2}=-N^{2}dt^{2}+a^{2}\left(  t\right)  \left(  dr^{2}+r^{2}\left(
d\theta^{2}+\sin^{2}\theta d\varphi^{2}\right)  \right)  . \label{ww.16}%
\end{equation}

We assume the comoving observer $u_{\mu}=\frac{1}{N}\delta_{\mu}^{t}$, and
$H=\frac{1}{N}\frac{\dot{a}}{a}$ is the Hubble constant. For the scalar field
$\phi$ we assume that it inherits the symmetries of the background space, such
that $\phi=\phi\left(  t\right)  .$ Hence, the gravitational field equations
are%
\begin{equation}
3H^{2}-\frac{\lambda}{2N^{2}}\dot{\phi}^{2}-V\left(  \phi\right)
-e^{-\frac{\phi}{2}}\rho_{m}=0, \label{ww.17}%
\end{equation}%
\begin{equation}
\dot{H}+H^{2}+\frac{1}{6}e^{-\frac{\phi}{2}}\left(  \rho_{m}+3p_{m}\right)
+\frac{1}{3}\left(  \frac{\lambda}{N^{2}}\dot{\phi}^{2}-V\left(  \phi\right)
\right)  =0, \label{ww.18}%
\end{equation}%
\begin{equation}
\ddot{\phi}+3H\dot{\phi}+V\left(  \phi\right)  +\frac{1}{2\lambda}%
e^{-\frac{\phi}{2}}\rho_{m}=0, \label{ww.19a}%
\end{equation}%
\begin{equation}
\dot{\rho}_{m}+3NH\left(  \rho_{m}+p_{m}\right)  -\rho_{m}\dot{\phi}=0.
\label{ww.20a}%
\end{equation}

For an ideal gas, i.e. $p_{m}=w_{m}\rho_{m}\,$, equation (\ref{ww.20a}) gives
$\rho_{m}=\rho_{m0}a^{-3\left(  w_{m}+1\right)  }e^{\phi}$, where $\rho_{m0}$
is an integration constant which describes the energy density of the matter
source at the present time.

If we replace $\rho_{m}$ in the field equations (\ref{ww.17})-(\ref{ww.19a})
it is easy to see that the rest of the field equations can be reproduced from
the variation of the singular Lagrangian function%
\begin{equation}
\mathcal{L}\left(  N,a,\dot{a},\phi,\dot{\phi}\right)  =\frac{1}{N}\left(
-3a\dot{a}^{2}+\frac{\lambda}{2}a^{3}\dot{\phi}^{2}\right)  -N\left(
a^{3}V\left(  \phi\right)  +\rho_{m0}e^{\frac{\phi}{2}}a^{w_{m}}\right)  .
\label{ee.06}%
\end{equation}
Lagrangian function (\ref{ee.06}) is a singular Lagrangian since
$\frac{\partial\mathcal{L}}{\partial\dot{N}}=0.$ The second-order differential
equations (\ref{ww.18}), (\ref{ww.19a}) are the Euler-Lagrange equations with
respect to the variables $a,$ and $\phi$, i.e. equations $\frac{d}{dt}%
\frac{\partial\mathcal{L}}{\partial\dot{a}}-\frac{\partial\mathcal{L}%
}{\partial a}=0$ ; $\frac{d}{dt}\frac{\partial\mathcal{L}}{\partial\dot{\phi}%
}-\frac{\partial\mathcal{L}}{\partial\phi}=0,$ respectively. Furthermore,
equation (\ref{ww.17}) follows from the variation of (\ref{ee.06}) with
respect to the lapse function $N,$ that is, $\frac{\partial\mathcal{L}%
}{\partial N}=0$. Equation (\ref{ww.17}) is nothing else than a constraint
equation which can be seen as the Hamiltonian for the autonomous dynamical
system (\ref{ww.18}), (\ref{ww.19a}) if without loss of generality we consider
$N\left(  t\right)  =N\left(  a\left(  t\right)  ,\phi\left(  t\right)
\right)  $.

In the latter case we apply the theory of symmetries of differential equations
for the determination of conservation laws for the field equations.
Specifically, we consider the application of Noether's first theorem for the
constraint of the scalar field potential $V\left(  \phi\right)  $, such that,
the dynamical system described by the point-like Lagrangian (\ref{ee.06})
admits Noether point symmetries. Moreover, with the use of Noether's second
theorem conservation laws can be constructed.

\section{Exact and analytic solutions from symmetry analysis}

\label{sec3}

Without loss of generality we consider $N\left(  t\right)  =1$. Hence for the
infinitesimal generator%
\begin{equation}
X=\Xi\left(  t,a,\phi\right)  \partial_{t}+\eta^{a}\left(  t,a,\phi\right)
\partial_{a}+\eta^{\phi}\left(  t,a,\phi\right)  \partial_{\phi},
\end{equation}
and the point-like Lagrangian (\ref{ee.06}); the symmetry condition
(\ref{tr.05}) gives that for nonzero potential function $V\left(  \phi\right)
$, the Noether point symmetries for the cosmological model of our
consideration WIG are, the vector field $X_{1}=\partial_{t}$ for arbitrary
potential, while for%
\begin{equation}
V\left(  \phi\right)  =V_{0}\exp\left(  -\frac{\phi}{w_{m}-1}\right)  ~,
\label{ee.06a}%
\end{equation}
there exists the additional Noether symmetry%
\begin{equation}
X_{2}=2t\partial_{t}+\frac{2}{3}a\partial_{a}+4\left(  w_{m}-1\right)
\partial_{\phi}.
\end{equation}

The exponential potential plays an important role in scalar field cosmology.
In the absence of matter it provides a scaling solution which can describes
the acceleration phase of the universe see for instance \cite{Copeland} and
references therein.

From Noether's second theorem we are able to construct the conservation laws.
For the vector field $X_{1}$ the conservation laws is the Hamiltonian function
$\mathcal{H}=\left(  \frac{\partial L}{\partial\dot{a}}\dot{a}+\frac{\partial
L}{\partial\dot{\phi}}\dot{\phi}-\mathcal{L}\right)  $, where from
(\ref{ww.17}) it follows $\mathcal{H}=0$. Moreover, for the vector field
$X_{2}$ we find the conservation law $I_{0}=2t\mathcal{H}-4a^{2}\dot
{a}+4\lambda a^{3}\left(  w_{m}-1\right)  \dot{\phi}$, that is,%
\begin{equation}
I_{0}=4a^{2}\dot{a}-4\lambda a^{3}\left(  w_{m}-1\right)  \dot{\phi}~.
\label{ee.07}%
\end{equation}

We proceed with the application of the symmetry vector and of the conservation
laws for the determination of exact and analytic solutions for the field equations.

\subsection{Exact solution}

For the cosmological model with the scalar field potential (\ref{ee.06a}),
from the vector field $X_{2}$ we define the invariant functions $U_{1}%
=at^{-\frac{1}{3}}$, $U_{2}=\phi-2\left(  w_{m}-1\right)  \ln t$. We assume
that the invariant functions are constant, that is,
\begin{equation}
a\left(  t\right)  =a_{0}t^{\frac{1}{3}}\text{, }\phi\left(  t\right)
=2\left(  w_{m}-1\right)  \ln t+\phi_{0}\text{~.}\label{ee.08}%
\end{equation}
where $a_{0}=U_{1}$ and $\phi_{0}=U_{2}$, in which without loss of generality
we consider $\phi_{0}=0$.

By replacing in the field equations (\ref{ww.17})-(\ref{ww.20a}) it follows
that (\ref{ee.08}) is an exact solution if and only if%
\begin{align}
V_{0} &  =\frac{1}{3}\frac{w_{m}-1}{w_{m}+1}\left(  6\lambda-1-12w_{m}%
\lambda+6\lambda w_{m}^{2}\right)  ,\\
\rho_{m0} &  =\frac{2}{3}\frac{a_{0}^{3\left(  1+w_{m}\right)  }}{3\left(
1+w_{m}\right)  }\left(  6\lambda-1+12w_{m}\lambda+6\lambda w_{m}^{2}\right)
~.
\end{align}

However, exact solution (\ref{ee.08}) describes a universe by a stiff fluid.
The equation of the state parameter for the effective fluid is $w_{eff}=1$.

We continue with the determination of the analytic solution for the given model.

\subsection{Hamilton-Jacobi equation\qquad}

In order to solve the field equations we apply the Hamilton-Jacobi approach.
We define the new variable $\psi=\phi-6\left(  w_{m}-1\right)  \ln a$, such
that we write the field equations on the normal coordinates.

In the new variables the point-like Lagrangian (\ref{ee.06}) reads,%
\begin{equation}
\mathcal{L}\left(  a,\dot{a},\psi,\dot{\psi}\right)  =3\left(  1-6\left(
w_{m}-1\right)  ^{2}\lambda\right)  a\dot{a}^{2}-6\left(  w_{m}-1\right)
\lambda a^{2}\dot{a}\dot{\psi}-\frac{1}{2}\lambda a^{3}\dot{\psi}^{2}%
+a^{-3}\left(  \rho_{m0}e^{\frac{\psi}{2}}+V_{0}e^{-\frac{\psi}{w_{m-1}}%
}\right)  . \label{ee.09}%
\end{equation}

We define the momentum $p_{a}=\frac{\partial\mathcal{L}}{\partial\dot{a}}%
$,~$p_{\psi}=\frac{\partial\mathcal{L}}{\partial\dot{\psi}}$, that is,
\begin{align}
p_{a}  &  =6\left(  1-6\left(  w_{m}-1\right)  ^{2}\lambda\right)  a\dot
{a}-6\left(  w_{m}-1\right)  \lambda a^{2}\dot{\psi}~,\\
p_{\psi}  &  =-6\left(  w_{m}-1\right)  \lambda a^{2}\dot{a}-\lambda a^{3}%
\dot{\psi}~,
\end{align}
from where it follows the Hamiltonian%
\begin{equation}
\mathcal{H}\left(  a,\psi,p_{a},p_{\psi}\right)  \equiv\frac{1}{12a}p_{a}%
^{2}-\frac{w_{m}-1}{a^{2}}p_{a}p_{\psi}+\frac{\left(  6\left(  w_{m}-1\right)
^{2}\lambda-1\right)  }{2\lambda a^{3}}p_{\psi}^{2}-a^{-3}\left(  \rho
_{m0}e^{\frac{\psi}{2}}+V_{0}e^{-\frac{\psi}{w_{m-1}}}\right)  =0.
\label{ee.10}%
\end{equation}

In the new variables the conservation law $I_{0}$ reads%
\begin{equation}
I_{0}=\frac{2}{3}ap_{a}. \label{ee.11}%
\end{equation}

In expression (\ref{ee.10}), we replace $p_{a}=\frac{\partial}{\partial
a}S\left(  a,\psi\right)  $, $p_{\phi}=\frac{\partial}{\partial\psi}S\left(
a,\psi\right)  $,\ from where we derive the Hamilton-Jacobi equation%
\begin{align}
0  &  =\frac{1}{12a}\left(  \frac{\partial}{\partial a}S\left(  a,\psi\right)
\right)  ^{2}-\frac{w_{m}-1}{a^{2}}\left(  \frac{\partial}{\partial a}S\left(
a,\psi\right)  \right)  \left(  \frac{\partial}{\partial\psi}S\left(
a,\psi\right)  \right) \label{ee.12}\\
&  +\frac{\left(  6\left(  w_{m}-1\right)  ^{2}\lambda-1\right)  }{2\lambda
a^{3}}\left(  \frac{\partial}{\partial\psi}S\left(  a,\psi\right)  \right)
^{2}-a^{-3}\left(  \rho_{m0}e^{\frac{\psi}{2}}+V_{0}e^{-\frac{\psi}{w_{m-1}}%
}\right)  ~.\nonumber
\end{align}
Moreover, from the conservation law (\ref{ee.11}) we write the constraint
equation
\begin{equation}
I_{0}=\frac{2}{3}a\frac{\partial}{\partial a}S\left(  a,\psi\right)  .
\label{ee.001}%
\end{equation}

By solving the Hamilton-Jacobi equation we determine the functional form for
the Action $S\left(  a,\psi\right)  $, which can be used to write the field
equations into an equivalent system of two first-order ordinary differential
equations%
\begin{equation}
\dot{a}=\frac{1-w_{m}}{a^{2}}\left(  \frac{\partial S}{\partial\psi}\right)
+\frac{1}{6a}\left(  \frac{\partial S}{\partial a}\right)  ~,\label{ee.14}%
\end{equation}%
\begin{equation}
\dot{\psi}=\frac{\left(  6\left(  w_{m}-1\right)  ^{2}\lambda-1\right)
}{\lambda a^{3}}\left(  \frac{\partial S}{\partial\psi}\right)  +\frac
{1-w_{m}}{a^{2}}\left(  \frac{\partial S}{\partial a}\right)  ~.\label{ee.15}%
\end{equation}
For a specific expression for the Action $S\left(  a,\psi\right)  ,$ the
dynamical system (\ref{ee.14}), (\ref{ee.15}) describes the reduced system for
the field equations provided by the Hamilton-Jacobi theory and the application
of a transformation. Its solution can be used to write the $a$ and $\phi$ of
the original system, as we shall see in the following lines.

Consequently, for $I_{0}=0$ and (\ref{ee.001}), the Action is given by the
closed-form expression%
\begin{equation}
S\left(  a,\psi\right)  =F\left(  \psi\right)
\end{equation}
with
\begin{equation}
\left(  \frac{\partial F\left(  \psi\right)  }{\partial\psi}\right)
^{2}=\frac{2\lambda\left(  V_{0}+\rho_{m0}e^{\frac{w_{m}+1}{w_{m}-1}\psi
}\right)  }{e^{\frac{\psi}{w_{m}-1}}\left(  6\lambda\left(  w_{m}-1\right)
^{2}-1\right)  }~.
\end{equation}

On the other hand for $I_{0}\neq0$ and equation (\ref{ee.001}) the Action is%
\begin{equation}
S\left(  a,\psi\right)  =\frac{3}{2}I_{0}\ln a+F\left(  \psi\right)  ,
\end{equation}
with%
\begin{equation}
\left(  \frac{\partial F\left(  \psi\right)  }{\partial\psi}\right)
=\frac{6I_{0}e^{\frac{\psi}{w_{m-1}}}\left(  w_{m}-1\right)  \lambda\pm
\sqrt{\kappa\left(  \Psi\right)  }}{4e^{\frac{\psi}{w_{m}-1}}\left(
6\lambda\left(  w_{m}-1\right)  ^{2}-1\right)  },
\end{equation}
where
\begin{equation}
\kappa\left(  \Psi\right)  =2e^{\frac{\psi}{w_{m-1}}}\left(  3e^{\frac{\psi
}{w_{m-1}}}I_{0}^{2}+16V_{0}\left(  6\lambda\left(  w_{m}-1\right)
^{2}-1\right)  +16e^{\frac{1}{2}\frac{w_{m}+1}{w_{m}-1}\psi}\left(
6\lambda\left(  w_{m}-1\right)  ^{2}-1\right)  \rho_{m0}\right)  .
\end{equation}

Let us focus on the case with $I_{0}=0$ and for simplicity let us assume
$w_{m}=0$. The Action $S\left(  a,\psi\right)  $ reads%
\begin{equation}
\left(  \frac{\partial F\left(  \psi\right)  }{\partial\psi}\right)
^{2}=\frac{2\lambda\left(  V_{0}e^{\psi}+\rho_{m0}e^{\frac{\psi}{2}}\right)
}{6\lambda-1}\text{.}%
\end{equation}
Thus, the reduced system is
\begin{equation}
a^{2}\dot{a}=\pm\sqrt{\frac{2\lambda\left(  V_{0}e^{\psi}+\rho_{m0}%
e^{\frac{\psi}{2}}\right)  }{6\lambda-1}},
\end{equation}%
\begin{equation}
\dot{\psi}=\pm\frac{1}{a^{3}}\sqrt{\frac{2\left(  6\lambda-1\right)  }%
{\lambda}\left(  V_{0}e^{\psi}+\rho_{m0}e^{\frac{\psi}{2}}\right)  }.
\end{equation}

Therefore, $\frac{d\psi}{da}=\frac{1}{a}\frac{\left(  6\lambda-1\right)
}{\lambda},$ or $\psi\left(  a\right)  =\frac{\left(  6\lambda-1\right)
}{\lambda}\ln a$. From the latter, it follows that the Hubble function is
expressed by the closed-form expression%
\begin{equation}
\left(  H\left(  a\right)  \right)  ^{2}=\frac{2\lambda}{6\lambda-1}\left(
V_{0}a^{-\frac{1}{\lambda}}+\rho_{m0}a^{-3-\frac{1}{2\lambda}}\right)
\text{.}%
\end{equation}
This is an analytic solution which describes an equivalent system of two ideal
gases non-interacting with constant equation of state parameters.

For the arbitrary parameter $w_{m}$, it follows $\frac{d\psi}{da}=\frac{1}%
{a}\frac{\left(  6\lambda\left(  w_{m}-1\right)  ^{2}-1\right)  }{\lambda}$,
in which the Hubble function is determined
\begin{equation}
\frac{H\left(  a\right)  ^{2}}{H_{0}^{2}}=\bar{\Omega}_{1}a^{p_{1}}%
+\bar{\Omega}_{2}a^{p_{2}}. \label{sd1}%
\end{equation}
with $\bar{\Omega}_{1}=\bar{\Omega}_{1}\left(  \lambda,w_{m},V_{0}\right)  $ ,
$\bar{\Omega}_{2}=\bar{\Omega}_{2}\left(  \lambda,w_{m},\rho_{m0}\right)  $
and indices $p_{1}=-6w_{m}+\frac{1}{w_{m}-1}\lambda$, $p_{2}=-3+3w_{m}\left(
w_{m}-2\right)  -\frac{1}{2\lambda}$.

\begin{figure}[ptb]
\centering\includegraphics[width=1\textwidth]{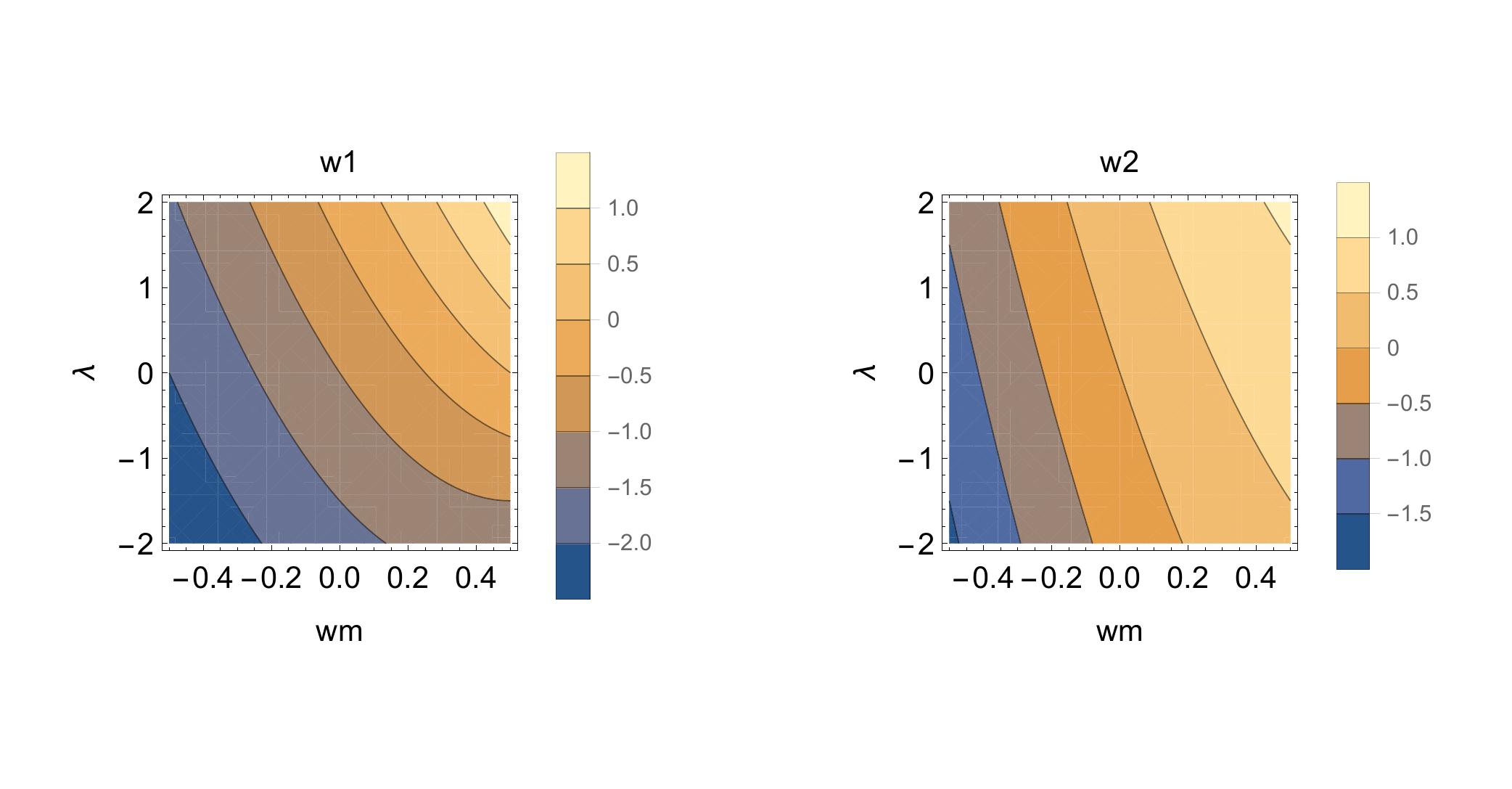}\caption{Contour plots
for the effective equation of state parameters $w_{1}\left(  w_{m}%
,\lambda\right)  =-1-\frac{1}{3}p_{1}\left(  w_{m},\lambda\right)  $
and\textbf{ }$w_{2}\left(  w_{m},\lambda\right)  =-1-\frac{1}{3}p_{2}\left(
w_{m},\lambda\right)  $ in the two-dimensional space of the free variables
$\left(  w_{m},\lambda\right)  $. }%
\label{fig1}%
\end{figure}

The exact solution (\ref{sd1}) describes a universe with two ideal gases
$\left(  \rho_{1},p_{1};p_{1}=w_{1}\rho_{1}\right)  $ , $\left(  \rho
_{2},p_{2};p_{2}=w_{2}\rho_{2}\right)  ~$ with equation of state parameters
$w_{1}\left(  w_{m},\lambda\right)  =-1-\frac{1}{3}p_{1}\left(  w_{m}%
,\lambda\right)  $ and $w_{2}\left(  w_{m},\lambda\right)  =-1-\frac{1}%
{3}p_{2}\left(  w_{m},\lambda\right)  $. In Fig. \ref{fig1} we present the
contour plots for the effective equation of state parameters $w_{1}\left(
w_{m},\lambda\right)  $ and\textbf{ }$w_{2}\left(  w_{m},\lambda\right)  ~$ in
the two-dimensional space of the free variables $\left(  w_{m},\lambda\right)
$. Such models have been widely studied before in the literature, see for
instance \cite{jm01} and reference therein. Indeed, such exact solution
provides physical parameters which can constrain some of the cosmological observations.

Indeed, for specific values of the free parameters $\left(  w_{m}%
,\lambda\right)  $ we can recover various eras of the cosmological evolution.
For the sets of parameters $\left(  p_{1}=0,p_{2}=-3\right)  $ or $\left(
p_{1}=-3,p_{2}=0\right)  $ the Hubble function for the $\Lambda$CDM is
recovered and the scale factor is that of $a\left(  t\right)  =a_{0}%
\sinh^{\frac{2}{3}}\left(  \frac{3}{2}H_{0}t\right)  $. On the other hand, for
$\left(  p_{1}=0,p_{2}=-4\right)  $ or $\left(  p_{1}=-4,p_{2}=0\right)  $ the
solution with a cosmological constant and a radiation fluid is provided.
Moreover, other dark energy models can be recovered where $w_{1}$ or $w_{2}$
take values lower than $-1$.

The deceleration parameter (\ref{sd1})\ can be easily derived as $q\left(
a\right)  =-1-\frac{d\ln H}{da}$ is derived.$~$From the constraints $\left(
p_{1}=0,p_{2}=-3\right)  $ or $\left(  p_{1}=-3,p_{2}=0\right)  $ we calculate
$\left(  w_{m},\lambda\right)  =\left(  -3,84\right)  $ and $\left(
w_{m},\lambda\right)  =\left(  0,0\right)  $. Thus in the present time, where
$a=1$, the deceleration parameter is calculated to be a function of the form,
$q\left(  \bar{\Omega}_{1}\right)  =3\bar{\Omega}_{1}-2$ and $q\left(
\bar{\Omega}_{1}\right)  =1-3\bar{\Omega}_{1}$. The present value of $q$ is
considered to be $q\simeq0.6~$\cite{jm01}, hence we derive $\bar{\Omega}%
_{1}\simeq0.86$ or $\bar{\Omega}_{1}\simeq0.13$. Which is clear that the
specific gravitational model recovers some of the recent cosmological parameters.

\section{Conclusions}

\label{sec5}

In this study we applied the symmetry analysis of differential equations;
specifically, the Noether symmetry conditions, for the constraint of the
unknown parameters of a cosmological model with an ideal gas in WIG. A scalar
field is introduced in the gravitational Action Integral as a result of Weyl
geometry. The scalar field is minimally coupled with gravity but interacts
with the matter source. The gravitational field equations admit a
minisuperspace description. There exists a point-like Lagrangian function
which produces the field equations under variation.

From Noether's two theorems we were able to constrain the unknown potential
function for the scalar field and we found that when it is exponential as it
is given by expression (\ref{ee.06a}) the field equations admit a second
conservation law. Consequently the field equations form an integrable
dynamical system and the Hamilton-Jacobi equation was explicitly solved. For a
special value of the second conservation law we were able to write the Hubble
function in a closed-form expression, where the cosmological solution consists
of two ideal gases.

In the case of a general exponential potential $V\left(  \phi\right)
=V_{0}\exp\left(  -\frac{\phi}{\kappa}\right)  $. The closed-form solution
follows%
\begin{equation}
a\left(  t\right)  =a_{0}t^{\frac{2+\kappa}{3\left(  1+w_{m}\right)  }}%
~,~\phi\left(  t\right)  =2\kappa\ln t~, \label{sd2}%
\end{equation}
with constraints for the free parameters of the model%
\begin{align}
V_{0}  &  =-\frac{\kappa}{3\left(  1+w_{m}\right)  ^{2}}\left(  6\kappa
\lambda\left(  w_{m}^{2}-1\right)  -2-\kappa\right)  ~,\\
\rho_{m0}  &  =-\frac{2a_{0}^{3\left(  1+w_{m}\right)  }}{3\left(
1+w_{m}\right)  ^{2}}\left(  6\kappa\lambda\left(  w_{m}^{2}+1\right)
-2-\kappa\right)  ~.
\end{align}

The latter solution is an exact solution of the model and not the general
solution, because it has fewer free parameters than the degrees of freedom of
the theory and it is valid only for a specific set of initial conditions for
the original dynamical system. However, it is clear that cosmology in WIG in
terms of the background space provides cosmological solutions of special interest.

At this point it is important to mention that the cosmological perturbations
for a model with background equations similar with that of the present study
were investigated before in \cite{ss1}. Thus similar results are expected and
for our model.

The scale factor (\ref{sd2}) describes a universe dominated by a perfect fluid
with constant equation of state parameter, $w=\frac{2w_{m}-\lambda}{2+\kappa}%
$. For $w<-\frac{1}{3}$, the exact solution describes an accelerated universe,
and this asymptotic solution can be used for the description of the early
acceleration phase of the universe \cite{dw1}. This study provides for the
first time analytic solutions for the Einstein-Weyl Integrable theory with a
nonconstant potential function and nonzero matter component. \ Analytic and
exact solutions are essential for the study of a gravitational theory because
we know that when a dynamical system is integrable there exist actual
solutions which can describe the dynamics. The concept of the integrability in
gravitational theories has been widely investigated by many scientists, see
for instance \cite{int1,int2,int3,int4}.

From this analysis\ we conclude that the Noether symmetry analysis is a
geometric criterion which provides physically accepted solutions in
cosmological studies.


\begin{thebibliography}{99}                                                                                               %


\bibitem {ns1}R. de Ritis, G. Marmo, G. Platania, C. Rubano, P. Scudellaro and
C. Stornaiolo, New approach to find exact solutions for cosmological models
with a scalar field, Phys. Rev. D 42, 1091 (1990)

\bibitem {ns2}K. Rosquist and C. Uggla, Killing tensors in two-dimensional
space-times with applications to cosmology. J. Math. Phys. 32, 3412 (1991)

\bibitem {ns3}S. Capozziello, P. Martin-Moruno, and C. Rubano, Dark energy and
dust matter phases from an exact f (R)-cosmology model, Phys. Lett. B 664, 12 (2008)

\bibitem {ns4}S. Cotsakis, P.G.L. Leach, C. Pantazi, Symmetries of homogeneous
cosmologies. Gravit. Cosmol. 4, 314 (1998)

\bibitem {ns5}N. Dimakis, T. Christodoulakis, and P.A. Terzis, FLRW metric f
(R) cosmology with a perfect fluid by generating integrals of motion. J. Geom.
Phys. 77, 97 (2012)

\bibitem {ns6}J. A. Belinchon, T. Harko and M.K. Mak, Exact Scalar-Tensor
Cosmological Solutions via Noether Symmetry. Astrophys. Space Sci. 361, 52 (2016)

\bibitem {ns7}S. Dutta and S. Chakraborty, A study of phantom scalar field
cosmology using Lie and Noether symmetries, Int. J. Mod. Phys 25, 1650051 (2016)

\bibitem {bas1}S. Basilakos, M. Tsamparlis and A. Paliathanasis, Using the
Noether symmetry approach to probe the nature of dark energy, Phys. Rev. D 86,
103512 (2011)

\bibitem {pl1}A. G. Riess, et al., Observational Evidence from Supernovae for
an Accelerating Universe and a Cosmological Constant, Astron J. 116, 1009 (1998)

\bibitem {pl2}S. Perlmutter, et al., Measurements of $\Omega$ and $\Lambda$
from 42 High-Redshift Supernovae, Astrophys. J. 517, 565 (1998)

\bibitem {pl3}P.~A.~R.~Ade et al. [Planck Collaboration], Planck 2015 results.
XIII. Cosmological parameters, Astron.\ Astrophys.\ 594, A13 (2016)

\bibitem {pl4}Planck Collaboration: N. Aghanim et al., Planck 2018 results.
VI. Cosmological parameters, Astron. Astrophy. 641, A6 (2020)

\bibitem {Ratra}B. Ratra and P.J.E. Peebles, Cosmological consequences of a
rolling homogeneous scalar field, Phys. Rev. D 37, 3406 (1988)

\bibitem {Barrow93}J.D. Barrow and P. Saich, Scalar-field cosmologies, Class.
Quant. Grav. 10, 279 (1993))

\bibitem {Copeland}E.J. Copeland, M. Sami and S. Tsujikawa, Dynamics of dark
energy, Int. J. Mod. Phys. D 15, 1753 (2006)

\bibitem {Overduin}J.M. Overduin and F.I. Cooperstock, Evolution of the scale
factor with a variable cosmological term, Phys. Rev. D 58, 043506 (1998)

\bibitem {Basil}S. Basilakos, M. Plionis and J. Sola, Hubble expansion and
structure formation in time varying vacuum models, Phys. Rev. D 80, 083511 (2009)

\bibitem {Buda}H.A. Buchdahl, Non-Linear Lagrangians and Cosmological Theory,
Mon. Not. Roy. Astron. Soc. 150, 1 (1970)

\bibitem {fq}F.K. Anagnostopoulos, S. Basilakos and E.N. Saridakis, First
evidence that non-metricity f(Q) gravity could challenge $\Lambda$CDM, Phys.
Lett. B 822, 136634 (2021)

\bibitem {Ferraro}R. Ferraro and F. Fiorini, Modified teleparallel gravity:
Inflation without an inflaton, Phys. Rev. D 75, 084031 (2007)

\bibitem {faraonibook}V. Faraoni, Cosmology in Scalar-Tensor Gravity,
Fundamental Theories of Physics vol. 139, (Kluwer Academic Press: Netherlands, 2004)

\bibitem {Ame10}L. Amendola and S. Tsujikawa, Dark Energy Theory and
Observations, (Cambridge University Press: Cambridge, 2010)

\bibitem {rev}E. Di Valentino, O. Mena, S. Pan, L.Visinelli et al., In the
realm of the Hubble tension---a review of solutions, Class. Quantum Grav. 38,
153001 (2021)

\bibitem {ind}S. Nojiri, S.D. Odintsov and V.K. Oikonomou, Modified Gravity
Theories on a Nutshell: Inflation, Bounce and Late-time Evolution,\ Phys.
Rept. 692, 1 (2017)

\bibitem {salim96}J.M. Salim and S.L. Saut\'{u}, Gravitational theory in Weyl
integrable spacetime, Class. Quantum Grav. 13, 353 (1996)

\bibitem {ww1}M. Yu Konstantinov and V.N. Melnikov, Integrable Weyl Geometry
in Multidimensional Cosmology. Numerical Investigation, Int. J. Mod. Phys.
D\ 4, 339 (1995)

\bibitem {ww2}C. Romero, J.B. Fonseca-Neto and M.L. Pucheu, General relativity
and Weyl geometry, Class. Quantum Grav. 29, 155015 (2012)

\bibitem {ww3}A. Paliathanasis and G. Leon, Integrability and cosmological
solutions in Einstein-\ae ther-Weyl theory, EPJC 81, 255 (2021)

\bibitem {ww4}R. Gannouji, H. Nandan and N. Dadhich, FLRW cosmology in
Weyl-Integrable Space-Time, JCAP 11, 051 (2021)

\bibitem {ww5}J. Miritzis, Acceleration in Weyl Integrable Spacetime, Int. J.
Mod. Phys. D 22, 1350019 (2013)

\bibitem {in2}W. Yang, S. Pan and A. Paliathanasis, Cosmological constraints
on an exponential interaction in the dark sector, MNRAS 482, 1007 (2019)

\bibitem {in4}W. Yang, A. Mukherjee, E. Di Valentino and S. Pan, Interacting
dark energy with time varying equation of state and the H0 tension,
Phys.\ Rev.\ D 98, 123527 (2018)

\bibitem {in5}S. Pan and G.S. Sharov, A model with interaction of dark
components and recent observational data, MNRAS 472, 4736 (2017)

\bibitem {gr}G. Panotopoulos and I. Lopes, Interacting dark sector: Lagrangian
formulation based on two canonical scalar fields, Phys.\ Rev. D 104, 083512 (2021)

\bibitem {lie1}S. Lie, Theorie der Transformationsgruppen I, Leipzig: B. G.
Teubner (1888)

\bibitem {lie2}S. Lie, Theorie der Transformationsgruppen II, Leipzig: B. G.
Teubner (1888)

\bibitem {lie3}S. Lie, Theorie der Transformationsgruppen III, Leipzig: B. G.
Teubner (1888)

\bibitem {noe18}E. Noether, Invariante Variationsprobleme
\textit{K\"{o}niglich Gesellschaft der Wis}senschaften G\"{o}ttingen
Nachrichten Mathematik-physik Klasse 2, 235-267 (1918)

\bibitem {h1}Hamel G, Ueber die Grundlagen der Mechanik, Mathematische Annalen
66, 350-397 (1908)

\bibitem {h2}G. Hamel, Ueber ein Prinzig der Befreiung bei Lagrange,
Jahresbericht der Deutschen Mathematiker-Vereinigung 25, 60-65 (1917)

\bibitem {h3}G. Herglotz, \"{U}ber den vom Standpunkt des
Relativit\"{a}tsprinzips aus als starr zu bezeichnenden K\"{o}rper, Annalen
der Physik 336, 393-415 (1910)

\bibitem {h5}A. Kneser, Kleinste Wirkung und Galileische Relativit\"{a}t
Mathematische Zeitschrift 2, 326-349 (1918)

\bibitem {h6}F. Klein, K\"{o}niglich Gesellschaft der Wissenschaften
G\"{o}ttingen Nachrichten Mathematik-physik Klasse 2 (1918)

\bibitem {nonle}A.K.\ Halder, A. Paliathanasis and P.G.L. Leach, Noether's
Theorem and Symmetry, Symmetry 10, 744 (2018)

\bibitem {anrv}M. Tsamparlis and A. Paliathanasis, Symmetries of Differential
Equations in Cosmology, Symmetry 10, 233 (2018)

\bibitem {ap1}T. Pailas, N. Dimakis, A. Paliathanasis, P.A.\ Terzis and
T.\ Christodoulakis, Infinite dimensional symmetry groups of the Friedmann
equations, Phys. Rev. D 102, 063524 (2020)

\bibitem {ap2}S. Dussault and V. Faraoni, A new symmetry of the spatially flat
Einstein--Friedmann equations, EPJC 80, 1002 (2020)

\bibitem {ap3}S. Dussault, V. Faraoni and A. Giusti, Analogies between
Logistic Equation and Relativistic Cosmology, Symmetry 13, 704 (2021)

\bibitem {ap4}G.Z. Abebe, S.D. Maharaj and K.S. Govinder, Gen. Rel. Grav. 46,
1733 (2014)

\bibitem {ss1}L. Amendola, Perturbations in a coupled scalar field cosmology,
MNRAS 312, 521 (2000)

\bibitem {jm01}A. Paliathanasis, M.\ Tsamparlis, S. Basilakos and J.D. Barrow,
Phys. Rev. D 93, 043528 (2016)

\bibitem {dw1}D. Wands, E.J. Copeland and A.R. Liddle, Ann. N.Y. Acad. Sci.
688, 647 (1993)

\bibitem {int1}M. Francaviglia and J. Kijowski, Gen. Relativ. Grav. 12, 279 (1980)

\bibitem {int2}E. Pozdeeva and S. Vernov, EPJ Web Conferences 125, 03008 (2016)

\bibitem {int3}V.R. Ivanov and S.Y.\ Vernov, EPJC 81, 985 (2021)\qquad

\bibitem {int4}V. R. Gavrilov, V. D. Ivashchuk and V. N. Melnikov, J. Math.
Phys. 36, 5829 (1995)
\end{thebibliography}
\end{document}